\documentstyle[12pt]{article}
\input epsfig.sty
\input psfig.sty
\def\nn{\noindent}
\def\Re{{\cal R \mskip-4mu \lower.1ex \hbox{\it e}\,}}
\def\Im{{\cal I \mskip-5mu \lower.1ex \hbox{\it m}\,}}
\def\gsim{\,\,\rlap{\raise 3pt\hbox{$>$}}{\lower 3pt\hbox{$\sim$}}\,\,}
\def\lsim{\,\,\rlap{\raise 3pt\hbox{$<$}}{\lower 3pt\hbox{$\sim$}}\,\,}
\def\gm2{(g - 2)_\mu}
\def\ie{{\it i.e.}}
\def\eg{{\it e.g.}}
\def\mpl{\ifmmode \overline M_{Pl}\else $\bar M_{Pl}$\fi}

\def\etal{{\it et al.}}
\def\ibid{{\it ibid}.}
\def\sub#1{_{\lower.25ex\hbox{$\scriptstyle#1$}}}
\def\tev{\,{\ifmmode\mathrm {TeV}\else TeV\fi}}
\def\gev{\,{\ifmmode\mathrm {GeV}\else GeV\fi}}
\def\mev{\,{\ifmmode\mathrm {MeV}\else MeV\fi}}
\def\to{\rightarrow}
\def\slash{\not\!}
\def\subw{_{\rm w}}
\def\mh{\ifmmode m\sbl H \else $m\sbl H$\fi}
\def\mch{\ifmmode m_{H^\pm} \else $m_{H^\pm}$\fi}
\def\mt{\ifmmode m_t\else $m_t$\fi}
\def\mc{\ifmmode m_c\else $m_c$\fi}
\def\mz{\ifmmode M_Z\else $M_Z$\fi}
\def\mw{\ifmmode M_W\else $M_W$\fi}
\def\mws{\ifmmode M_W^2 \else $M_W^2$\fi}
\def\mhs{\ifmmode m_H^2 \else $m_H^2$\fi}   
\def\mzs{\ifmmode M_Z^2 \else $M_Z^2$\fi}
\def\mts{\ifmmode m_t^2 \else $m_t^2$\fi}
\def\mcs{\ifmmode m_c^2 \else $m_c^2$\fi}
\def\mchs{\ifmmode m_{H^\pm}^2 \else $m_{H^\pm}^2$\fi}
\def\ztwo{\ifmmode Z_2\else $Z_2$\fi}
\def\zone{\ifmmode Z_1\else $Z_1$\fi}
\def\mtwo{\ifmmode M_2\else $M_2$\fi}
\def\mone{\ifmmode M_1\else $M_1$\fi}
\def\tb{\ifmmode \tan\beta \else $\tan\beta$\fi}
\def\xw{\ifmmode x\subw\else $x\subw$\fi}
\def\ch{\ifmmode H^\pm \else $H^\pm$\fi}
\def\lum{\ifmmode {\cal L}\else ${\cal L}$\fi}
\def\inpb{\,{\ifmmode {\mathrm {pb}}^{-1}\else ${\mathrm {pb}}^{-1}$\fi}}
\def\infb{\,{\ifmmode {\mathrm {fb}}^{-1}\else ${\mathrm {fb}}^{-1}$\fi}}
\def\epem{\ifmmode e^+e^-\else $e^+e^-$\fi}
\def\ppb{\ifmmode \bar pp\else $\bar pp$\fi}
\def\bsg{\ifmmode B\to X_s\gamma\else $B\to X_s\gamma$\fi}
\def\bsll{\ifmmode B\to X_s\ell^+\ell^-\else $B\to X_s\ell^+\ell^-$\fi}
\def\bstt{\ifmmode B\to X_s\tau^+\tau^-\else $B\to X_s\tau^+\tau^-$\fi}
\def\lamt{\ifmmode \tilde\lambda\else $\tilde\lambda$\fi}
\def\shat{\ifmmode \hat s\else $\hat s$\fi}
\def\that{\ifmmode \hat t\else $\hat t$\fi}
\def\uhat{\ifmmode \hat u\else $\hat u$\fi}

\newskip\zatskip \zatskip=0pt plus0pt minus0pt
\def\matth{\mathsurround=0pt}
\def\lsim{\mathrel{\mathpalette\atversim<}}
\def\gsim{\mathrel{\mathpalette\atversim>}}
\def\atversim#1#2{\lower0.7ex\vbox{\baselineskip\zatskip\lineskip\zatskip
  \lineskiplimit 0pt\ialign{$\matth#1\hfil##\hfil$\crcr#2\crcr\sim\crcr}}}

\renewcommand{\thefootnote}{\fnsymbol{footnote}}

\begin{document} \begin{titlepage} 
\rightline{\vbox{\halign{&#\hfil\cr
&SLAC-PUB-8469\cr
&June 2000\cr}}}
\begin{center}

{\Large\bf The $g-2$ of the Muon in Localized Gravity Models}
\footnote{Work supported by the Department of 
Energy, Contract DE-AC03-76SF00515}
\medskip

\normalsize 
{\large H. Davoudiasl, J.L. Hewett, and T.G. Rizzo } \\
\vskip .3cm
Stanford Linear Accelerator Center \\
Stanford University \\
Stanford CA 94309, USA\\
\vskip .3cm

\end{center}

\begin{abstract} 
The $(g-2)$ of the muon is well known to be an important model building 
constraint on theories beyond the Standard Model.  In this paper, we examine 
the contributions to $(g-2)_\mu$ arising in the Randall-Sundrum model of 
localized gravity for the case where the Standard Model gauge fields and 
fermions are both in the bulk. Using the current experimental world average 
measurement for $(g-2)_\mu$, we find that strong constraints can be placed 
on the mass of the lightest gauge Kaluza-Klein excitation for a narrow 
part of the allowed range of the assumed universal 5-dimensional fermion mass 
parameter, $\nu$. However, employing both perturbativity and fine-tuning 
constraints we find that we can further restrict the allowed range of 
the parameter $\nu$ to only one fourth of its previous size. The scenario with 
the SM in the RS bulk is thus tightly constrained, being viable for only a 
small region of the parameter space. 
\end{abstract} 




\renewcommand{\thefootnote}{\arabic{footnote}} \end{titlepage}


The existence of extra spacetime dimensions has recently been
suggested\cite{nima,rs,old} as a means to explain the hierarchy.
In one scenario of this kind from Arkani-Hamed, Dimopoulos, 
and Dvali (ADD)\cite{nima}, the apparent hierarchy is generated by a large 
volume for the extra dimensions.  In this case, the fundamental Planck scale in 
$4+n$-dimensions, $M$, can be reduced to the TeV scale and is related 
to the observed 4-d Planck scale, $\mpl$, through the volume $V_n$ 
of the compactified 
dimensions, $\mpl^2=V_nM^{2+n}$.  In a second scenario due to
Randall and Sundrum (RS)\cite{rs}, the observed hierarchy is induced through 
an exponential warp factor which arises from a non-factorizable geometry.  
An exciting feature of these approaches is that they both lead to concrete and 
distinctive phenomenological tests\cite{pheno,dhr1} at the TeV scale.  

In addition to collider tests, loop-order processes, such as rare
transitions which are suppressed in the Standard Model (SM) or
radiative corrections to perturbatively calculable processes, can provide 
complementary information about new physics.  One such traditional quantity is 
the $(g-2)$ of the $\mu${\cite {gm2}}. Currently the SM prediction is 
approximately $1\sigma$ higher than that of the World Average measured 
value, with
the difference between the theoretical and experimental results being
$a^{exp}_\mu-a^{SM}_\mu=(43\pm 45)\times 10^{-10}$, where 
$a=(g-2)/2$. This corresponds to a $95\%$ CL upper bound on the magnitude of 
a new negative 
contribution, $\Delta a_\mu$, of $-3.1\times 10^{-9}$. The E821 experiment at 
BNL is expected to reduce the experimental error on $a_\mu^{exp}$ by 
approximately an order of magnitude during the next few years to 
the level of 0.35ppm which is below the current SM theory error of 0.60ppm. 
The SM error will also decrease in the future as more data on the $R$ ratio 
in the low energy region becomes available. 
The size of the contribution to $(g-2)_\mu$ in the ADD scenario has been 
calculated in Ref.{\cite {gra}} and results in interesting constraints. In 
this paper, we examine this quantity within the RS model in the case where 
the SM 
fields propagate in the bulk with the expectation from our earlier 
work{\cite {dhr}} that existing data will yield 
interesting bounds over a region of the parameter space.
With an anticipated ten-fold increase in the experimental 
precision in the not too distant future, these 
bounds should soon improve if no signal for new physics is observed. 

In its original construction, the RS model consists of two 3-branes each being
stabilized{\cite {gw2}} at an $S^1/Z_2$ orbifold fixed point with a 
separation of $\pi r_c$ between the branes in an additional dimension 
denoted as $r_c\phi$. The model initially
postulated that only gravity was allowed to propagate in the 
higher dimensional anti-deSitter bulk with the SM fields being confined to 
one of the 3-branes. The exponential `warp' factor $e^{-kr_c\phi}$, with $k$ 
being a 5-d space-time curvature parameter of order the Planck scale, 
is responsible for generating the observed hierarchy assuming the
scale of physics on the SM brane located at $\phi=\pi$ is 
$\Lambda_\pi=\mpl e^{-kr_c\pi}\simeq 1$ TeV
with $kr_c\sim 11-12$. The usual 4-d Planck scale and that of the 
original 5-d theory are found to be related via $\mpl^2=M_5^3/k$. 
Recently, a series of authors{\cite {dhr,others}} have 
considered peeling the SM 
gauge and matter fields off of the wall in the limit where their back-reaction 
on the RS metric can be ignored. (There are a number of arguments which 
strongly suggest that if the Higgs is the source of electroweak symmetry 
breaking it must remain on the wall{\cite {dhr,others}}.) It is the existence 
of these SM bulk fields that allows for a potentially sizeable contribution to 
$(g-2)_\mu$. In 
what follows we use the notation as defined in the last paper listed in 
Ref.{\cite {dhr}}. 

When the SM gauge and matter fields are allowed to propagate in the bulk, 
there are three 
parameters that need to be specified to determine the phenomenological 
predictions of the RS model: $c\equiv k/\mpl$ 
which is expected to lie in the range 
0.01 to 1, the common dimensionless bulk mass parameter for the fermions
$\nu\equiv m/k$, where $m$ represents the 5-d fermion mass,
which is expected to be of order unity, and the 
mass of the lightest gauge, fermion or graviton Kaluza-Klein(KK) excitation. 
We remind the reader that a common value of $\nu$ for 
all fermions is not a necessary assumption but is certainly the 
simplest choice and the one which naturally avoids constraints associated 
with flavor changing neutral currents. For a fixed value 
of $\nu$,the entire KK spectrum is determined for all fields 
once the mass of a single KK excitation is known.  We recall that the KK
spectrums for gravitons, fermions, and gauge bosons are related\cite{dhr} by 
the roots of various Bessel functions and that 
all gauge bosons, \ie, gluons, $W$'s, $Z$'s and $\gamma$'s, have essentially 
the same excitation spectra.
  
Note that the limits we obtain below are derived under the 
assumption that no other new physics is present beyond what is considered
here.  As with all bounds obtained via indirect means, the presence
of additional new interactions may cancel the loop effects and erase or ease 
the constraints.

Consider the situation where we have two fermions in the bulk, $D$ and $S$, 
which have the quantum numbers of an $SU(2)_L$ doublet and singlet with weak 
hypercharges $Y=-1/2$ and $-1$, respectively. Following the notation of our 
previous work, their interactions with the gauge fields can be described by 
the action, 
\begin{equation}
S_{fV} =\int d^4x \int r_c~d\phi \sqrt{G} \left[V^M_n\left
({i\over {2}}\overline{S} \, \gamma^n\,  {\cal D}_M S + h. c.\right) - 
sgn(\phi) m_S \overline{S} S +(S\to D\right)],
\label{SfV}
\end{equation}
where, $G$ is the determinant of the metric tensor, $V$ is the vielbein, 
${\cal D}_M$ is a covariant derivative and $h. c.$ 
denotes the Hermitian conjugate term. Here, as discussed above, we will assume 
that $m_D=m_S=k\nu$. Note that gauge interactions do not mix the $D$ and $S$ 
fields.  The $D$ and $S$ fields also interact with 
the Higgs isodoublet field(s), $H^0$, which reside on the wall, \ie, 
\begin{equation}
S_{fH} ={\lambda \over {k}}\int d^4x \int d\phi \sqrt{G} \ \overline{S}DH^0 \ 
\delta(\phi-\pi)+ h. c.,
\label{SfH}
\end{equation}
with $\lambda$ being a dimensionless Yukawa coupling. 
Due to the KK mechanism the fields $D_{L,R}^{(n)}$ and $S_{L,R}^{(n)}$ 
form separate 4-d towers of Dirac fermions which are degenerate level by 
level. The KK expansion can be written as 
$D=\sum D_L^{(n)}(x)\chi^{(n)}(\phi)+ D_R^{(n)}(x)\tau^{(n)}(\phi)$ and 
$S=\sum S_L^{(n)}(x)\tau^{(n)}(\phi)+ S_R^{(n)}(x)\chi^{(n)}(\phi)$ where the 
$\nu$-dependent 
$\chi(\tau)$ fields are $Z_2$ even(odd) and are given explicitly in our 
previous paper. Note that the $Z_2$ orbifold 
symmetry allows couplings of the type 
$\overline{D_L}^{(n)}S_R^{(m)}+h. c.$ but {\it not} ones of the form 
$\overline{D_R}^{(n)}S_L^{(m)}+h. c.$ since the $Z_2$ 
odd wavefunctions vanish on both of the boundaries. After shifting the Higgs 
field $H^0 \to e^{kr_c\pi}H$ so 
that it is canonically normalized, the value of $\lambda$ is fixed as a 
function of $\nu$ by the   requirement 
that the coupling of the $D_L$ and $S_R$ zero 
modes obtains a mass, $m_\mu$, once the Higgs gets a vev, 
$v_4=v_{SM}/\sqrt {2}$ with $v_{SM}\simeq 246$ GeV. (It is thus important to 
observe that in a theory with a fixed value of $\nu$ the set of Yukawa 
couplings $\lambda_f$ associated with the set of SM fermions is clearly 
hierarchical.)
This then fixes the couplings between a Higgs, a zero mode fermion and any 
tower member to be 
$C_{0i}^{ffH}=(-1)^i~{m_\mu \over {v_4}}\sqrt F$, as well as the 
coupling between two different tower members and the Higgs as 
$C_{ij}^{ffH}=(-1)^{i+j}~{m_\mu \over {v_4}}F$  
(up to a possible $\nu$-dependent sign where), 
\begin{equation}
F=2\Big|{1-\epsilon^{2\nu +1}\over {1+2\nu}}\Big|,
\end{equation}
with $\epsilon=e^{-kr_c\pi}\sim 10^{-16}$. Note that for negative values 
of $\nu$, the factor $F$ grows exponentially large.  

In terms of the $D$ 
and $S$ fields, the operator which generates the anomalous magnetic dipole 
moment of the $\mu$ can be written as $D_L^{(0)}\sigma_{\mu\nu}S_R^{(0)}
+h.c.$. This reminds us that this operator and the muon mass generating term 
have the same isospin and helicity structure such that a Higgs interaction is 
required in the form of a mass insertion to connect the two otherwise 
decoupled zero 
modes. We can think of this mass insertion as the interaction of a fermion 
with an external Higgs field that has been replaced by its vev.

\vspace*{-0.5cm}
\nn
\begin{figure}[htbp]
\centerline{
\psfig{figure=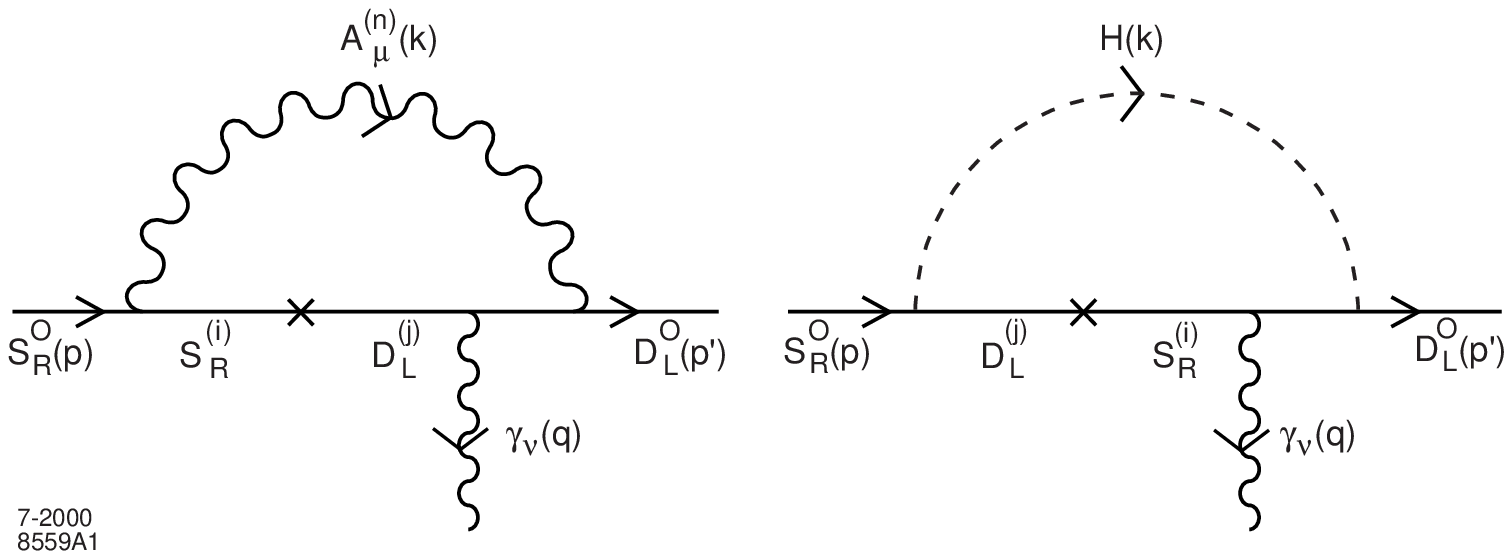,height=8.0cm,width=16.2cm,angle=0}}
\vspace*{0.1cm}
\caption[*]{Typical Feynman diagrams for the gauge and Higgs boson 
contribution to $(g-2)_\mu$. The mass insertion is denoted by the cross.}
\label{fig1}
\end{figure}
\vspace*{0.7mm}

Helicity flips play an 
important role in evaluating the contributions to $(g-2)_\mu$  
since muon KK excitations are now propagating inside the loop. 
As is well-known, for non-chiral couplings the contribution to the 
anomalous magnetic moment of a light fermion can be enhanced when a heavy 
fermion of mass $m_h$ participates inside the loop{\cite {stan}}.
There are a number of diagrams that can contribute to $(g-2)_\mu$ at one loop 
of which two are shown in Fig.~1. The diagram on the left corresponds to the 
exchange of a tower of the 4-d neutral gauge bosons, $\gamma^{(n)}$ and/or 
$Z^{(n)}$, which we will now discuss in detail. Due to gauge 
invariance we are free to choose a
particular gauge in order to simplify the calculation. Here, we make use of 
the $\xi=1$ unitary gauge where the 4-d propagator is just the flat space 
metric tensor{\cite {dpq}}. 
Hence, the loops with the 4-d components of the 
gauge fields and the ones with the fifth component need to be considered 
separately. In this example, the mass insertion takes place inside the loop 
before the photon is emitted. Clearly there are three other diagrams of this 
class: two with the mass insertion on an external leg and the third with the 
mass insertion inside the loop but after the photon is emitted. The amplitude 
arising from this vector exchange graph is given by 
\begin{eqnarray}
{\cal M}_V &=& C_LC_RC_{0in}^{ffA}C_{0jn}^{ffA}~\bar u(p')\ (-ie\gamma_\mu)P_L 
i{{\hat {\slash p'}+m_j} \over {\hat p'^2-m_j^2}} \ (-ie\gamma_\nu)P_L \, 
\nonumber \\  
 &\times &  i{{\hat {\slash p}+m_j} \over {\hat p^2-m_j^2}} \ (im_{ij}P_R) 
\ i{{\hat {\slash p}+m_i} \over {\hat p^2-m_i^2}} \ (-ie\gamma^\mu)P_R \ u(p) 
\ {-i\over {k^2-m_A^{(n)~2}}} +h. c. \,,
\end{eqnarray}
where $C_{L,R}$ are the corresponding couplings of the SM gauge boson 
to the $\mu$ in units of $e$ and $\hat p(\hat p')=p(p')-k$. The coefficients 
$C_{0in}^{ffA}$ are the reduced couplings between a zero mode fermion, a 
fermion tower member of mass $m_i$ and the $n^{th}$ gauge boson tower member. 
Here $m_i$ and $m_j$ are the masses of the $D$ or $S$ 
fermionic KK states and 
$m_A^{(n)}$ are the masses of the KK gauge tower states. 
Note that the mass insertion, $m_{ij}$, comes with a chirality factor that can 
be determined from the action $S_{fH}$; numerically, $m_{ij}=C_{ij}^{ffH}v_4$.
The amplitude where the mass insertion comes after the photon emission can 
be easily obtained by interchanging $i$ and $j$ in the resulting final 
amplitude expression. When the mass insertion 
occurs on an external leg it connects a zero mode with a tower mode and is 
given by $m_{0i}=C_{0i}^{ffH}v_4$. With some algebra it is straightforward to 
show that the corresponding amplitudes obtained in the two cases with 
external insertions are 
suppressed in comparison to the case of internal insertion 
by a factor of order $\sim m_\mu^2/M_{KK}^2$, 
where $M_{KK}$ is a typical large KK mass. 
In the case of the $W$ gauge boson 
tower graphs, since the $W$ couples only to the $D$'s, 
the mass insertion must occur on the incoming leg of the graph and the photon 
is emitted from the $W$; this graph can also 
be shown to produce a sub-leading contribution by a factor of 
order $\sim m_\mu^2/M_{KK}^2$. 
Thus, $W$ tower graphs can be safely ignored in comparison to those arising 
from the $Z$ and $\gamma$ towers and the resulting contribution from all 
of the 4-d vector exchanges, neglecting the subleading terms, is given by
\begin{eqnarray}
\Delta a_\mu^A &=& \sum_{i,j,k} 
{-3C_V^2K\alpha m_{ij}m_\mu \over {\pi}} \times \, \nonumber \\  
 & & \int_0^1~dx~\int_0^{1-x}~dy~\Bigg[{x(1-x-y)
\over {m_j^2(1-x-y)+m_i^2 y+m_A^{(k)2}x}} \ +(i\to j) \Bigg] \,, 
\end{eqnarray}
where the sum is over all internal KK states, the $i \to j$ represents the 
addition of the other internal insertion graph and 
$C_V^2=C_{0ik}^{ffA}C_{0jk}^{ffA}$ with $K$ given by 
\begin{equation}
K=\Bigg[1+{{(1-4x_w)^2-1}\over {16x_w(1-x_w)}}\Bigg]\,,
\end{equation}
where $x_w=\sin^2 \theta_w \simeq 0.2315$. Note that since the coefficients 
$C_{0ik}^{ffA}$ behave as $\sim 1/\sqrt F${\cite {dhr}} and $m_{ij}\sim F$ 
we may expect that the $\nu$ dependence of $\Delta a_\mu^A$ to be rather weak. 
In principle the sum extends over all of the internal KK states but in 
practice we find that truncating the sum after the first 20-40 members of each 
tower leads to a rather stable result. 

The next class of graphs is similar to the 4-d vector exchange, but in the 
$\xi=1$ gauge, now involves the fifth component of the original 5-d field. 
Here it is important to recall that these fifth components are $Z_2$ odd 
fields thus 
connecting $S_L(D_L)$ with $S_R(D_R)$. The action $S_{fH}$ in Eq.(2) 
demonstrates that the Higgs boson 
does not interact with odd fields since they vanish on the wall.  
From this observation we conclude that in the case of fifth component vector 
exchanges the mass 
insertions can only occur on the external legs. By following similar 
algebraic manipulations as before it is easy to show that all 
of these contributions are always subleading by factors of order 
$\sim m_\mu^2/M_{KK}^2$ and thus their contribution to $(g-2)_\mu$ can be 
safely neglected.

Next, we turn to the possibility of Higgs exchange, also shown in Fig.~1. 
Ordinarily, one might dismiss such contributions as being small but they now 
involve the off-diagonal Higgs couplings discussed above which contain powers 
of the factor $F$ which grows large rapidly as $\nu$ grows negative. 
The amplitude 
for the Higgs graph shown in the figure is given by 
\begin{eqnarray}
{\cal M}_H &=& C_{0i}^{ffH}C_{0j}^{ffH}~\bar u(p')\ iP_R \ 
i{{\hat {\slash p'}+m_i} \over {\hat p'^2-m_i^2}} (-ie\gamma_\nu)P_R \ 
i{{\hat {\slash p}+m_i} \over {\hat p^2-m_i^2}} \, \nonumber \\  
 &\times &   (im_{ij}P_L) \ i{{\hat{\slash p}+m_j} \over {\hat p^2-m_j^2}} 
\ iP_R \ u(p) \ {i\over {k^2-m_H^2}} \ +h. c. \,,
\end{eqnarray}
using the notation above. Here the coefficients $C_{0i}^{FFH}$ are the reduced 
couplings of a Higgs boson to a zero mode fermion and an $i^{th}$ fermion 
tower member. 
The amplitude where the insertion and emission occur 
with the opposite order can be obtained in a straightforward manner and, as 
we now expect, the two diagrams with external insertions can be shown to be 
subleading. Combining the two dominant 
Higgs amplitudes we find the following contribution to $\Delta a_\mu$:
\begin{equation}
\Delta a_\mu^H=\sum_{i,j}  {C_H^2m_{ij}m_\mu \over {16\pi^2}} \   
\int_0^1~dx~\int_0^{1-x}~dy~
\Bigg[{{(1-x-y)(3y-3x-1)}\over {m_i^2(1-x-y)+m_j^2 y+m_H^2x}} \ 
+(i\to j)\Bigg]\,,
\end{equation}
where $C_H^2=C_{0i}^{ffH}C_{0j}^{ffH}$ with $m_H$ being the Higgs boson mass. 
The $i\to j$ term results from the addition of the other internal emission 
graph. 
(In our numerical analysis below we assume $m_H=120$ GeV; these results are not 
very sensitive to this particular choice.) Since $C_{0i}^{ffH}$ 
behave as $\sim \sqrt F$ and $m_{ij}\sim F$ we expect $\Delta a_\mu^H$ to have 
a strong $\nu$ dependence and to grow very rapidly as $\nu$ becomes 
increasingly negative. As in the vector case, truncating the sum over the KK 
fermion contributions after the first 20-40 
tower members have been included yields a numerically stable 
result.

What are the other potential contributions to $(g-2)_\mu$ in the RS model? 
The radion is the zero mode remnant scalar resulting from 
the KK decomposition of the 5-d graviton field. Since it couples diagonally to 
KK tower members, as does the zero mode graviton and photon, the $\mu$ 
continues through the diagram and no KK modes are excited. Loops involving 
radions{\cite {rad}} are thus easily 
shown to be small{\cite {last}} since both the radion couplings and the mass 
insertions in this case are not accompanied by any compensating powers of $F$. 
These contributions can be safely neglected.

The last remaining potential contribution arises from graviton loops which may 
be calculated via the Feynman rules given in {\cite {hanandjim}} with small 
modifications due to the fact that the SM fields are now in the bulk and have 
nontrival $Z_2$ parity. These diagrams lead to amplitudes which are found 
to be log-divergent and lead to cutoff ($\Lambda$) dependent results and 
are thus not well-defined.   The divergences, which occur in all graviton
diagrams, arise due to the fact that the operator describing, \eg, the 
fermion-fermion-graviton interaction is dimension-five and involves an
additional power of fermion  momentum.

This differs significantly from the results obtained by Graesser \cite{gra} 
in the case of the ADD model, where it was found that the total contribution
to $(g-2)_\mu$ due to gravity is finite.  This difference arises from a number of
sources: (i) in the ADD model the SM fields lie on the wall and only gravitons
are allowed to propagate in the bulk, whereas in the version of the RS model
under consideration here, both the SM gauge fields and fermions propagate in
the bulk.  (ii) In the ADD case, the graviton couplings are universal for all
KK tower members whereas in the RS scenario the couplings are 
KK excitation state  dependent
and also differ for fermions and gauge bosons for arbitrary values of $\nu$.
(iii) In the ADD case, each of the 5 diagrams shown in Fig. 1 of Graesser 
\cite{gra} was found to be log divergent with their sum, however, being finite,
since the divergences cancel at each KK level.  In the RS case, these 
cancellations cannot occur due to both the breakdown in universality of the
graviton couplings and the fact that the complete calculation of each diagram
involves different coupling coefficients and different numbers of KK states.
For example, in the diagram where the graviton is emitted off the fermion
KK line, we must sum over the triple product of coefficients $C_{0ik}^{ffG}
C_{0jk}^{ffG}m_{ij}$.  On the otherhand, 
the diagram involving the gauge-gauge graviton
vertex we must instead sum over the quartic product of coefficients
$C_{0ik}^{ffA}C_{0k\ell}^{AAG}C_{0n\ell}^{ffG}m_{in}$, where the sum extends over 
the KK towers for two
fermions, one graviton, and one photon.  Since the evaluation of
the coefficients involve $\nu$-dependent integrals over Bessel functions, it
is highly unlikely that the divergences encountered in each class of
diagrams can sum to zero unless a theorem demands that it
be so.  Thus, we expect that the complete contribution due to gravity to
remain log divergent when all diagrams are summed.

We can, however, make an estimate of the size of these graviton
contributions. As in the case of vector and Higgs boson exchange 
we expect 
graphs with internal insertions proportional to $F$ to dominate. Since the 
relevant vertex couplings $C_{0ik}^{ffG}$ scale as 
$\sim 1/\sqrt F${\cite {dhr}} we do not 
expect the graviton contribution to be strongly $\nu$ dependent as was also the 
case for vector exchange. An order of magnitude estimate suggests that 
\begin{equation}
\Delta a_\mu^G \simeq \sum_{ijk}  {C_{0ik}^{ffG}C_{0jk}^{ffG}m_{ij}m_\mu 
\over {16
\pi^2 \Lambda_\pi^2}} \ log \Bigg [{\Lambda^2\over {m_{KK_k}^2}}\Bigg]  \,,
\end{equation}
with the cutoff $\Lambda$ expected to be of order $\Lambda_\pi$.  The
origin of the various terms in this estimate are easy to identify: the
$C$'s are vertex functions, the $(16\pi^2)^{-1}$ is a typical loop factor,
the $\Lambda_\pi^{-1}$ appears in all graviton couplings to SM fields,
and the $m_{ij}m_\mu$ arises in the usual single mass insertion
approximation as seen above.  The $\log\Lambda^2$ is the divergence
discussed above and $m_{KK_i}$ represents canonical KK-tower masses
which appear in the loop.  Note that $\Lambda_\pi\to\infty$ as
$m_{KK_i}\to\infty$ and thus $\Delta a_\mu^G\to 0$ as the KK mass gets large.  
Inserting 
typical values of the parameters we estimate that $\Delta a_\mu^G$ should 
certainly be less than $10^{-(9-10)}$ and thus is at most comparable to the 
vector boson contribution.  (An explicit calculation of the diagram
containing the $\gamma ffG$ vertex confirms this expectation.)
As we will see this means that the graviton 
exchange contributions will then have very little effect upon our results.

\vspace*{-0.5cm}
\nn
\begin{figure}[htbp]
\centerline{
\psfig{figure=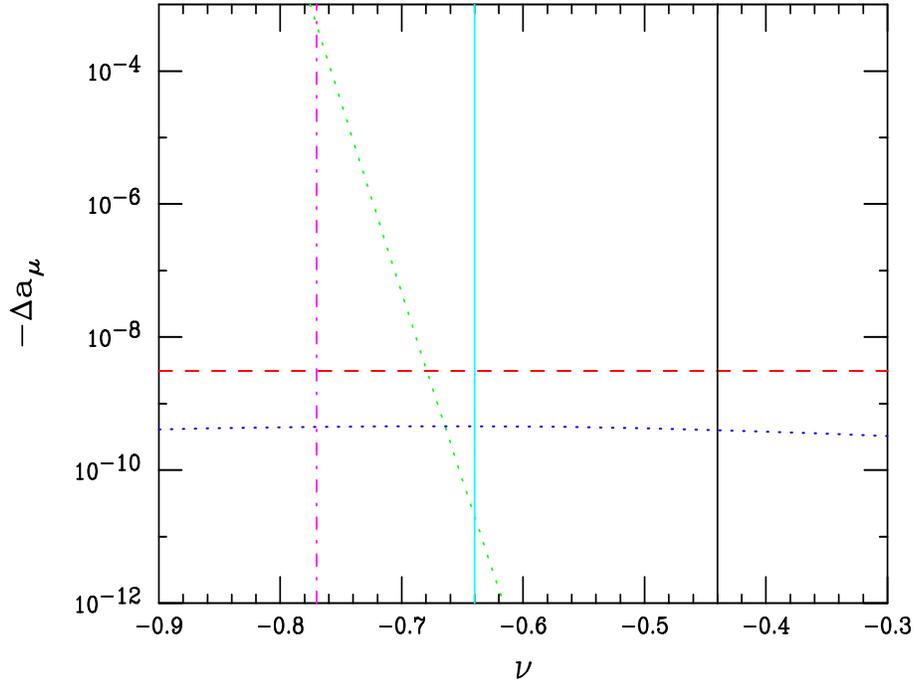,height=9.0cm,width=12cm,angle=90}}
\vspace*{0.1cm}
\caption[*]{Bounds on the parameter $\nu$ from considerations of $(g-2)_\mu$. 
The horizontal dotted curve shows the value of $-\Delta a_\mu^A$ while 
the steeply rising dotted curve is that for $-\Delta a_\mu^H$; both of 
which are calculated assuming the mass of the first KK gauge state to be 
1 TeV. The vertical 
dash dotted line arises from the Yukawa coupling perturbativity constraint 
discussed in the text while the left(right)solid line arises from the 
corresponding $\mu$(top)  
no fine-tuning mass constraint. Regions to the left of these lines are 
excluded. The horizontal dashed line is the current experimental upper bound at 
$95\%$ CL.}
\end{figure}
\vspace*{0.4mm}

Let us now turn to our numerical results which are summarized in Fig.~2. We 
first remind the reader that for 
a fixed value of $\nu$ specifying the mass of any single KK tower state 
determines the entire KK spectrum for fermions, gauge bosons and gravitons. 
Here we take the mass of the 
lightest gauge KK state to be 1 TeV with the results for both the vector and 
Higgs contributions scaling as 
$\sim (1~{\rm TeV}/M_{KK})^2$. In our previous work{\cite {dhr}} 
we have shown that for 
values of $\nu \gsim -0.3$ the masses of the KK states as well 
as $\Lambda_\pi$ are required to be in the multi-TeV range, disfavoring 
this model as a solution to the hierarchy problem. 
Also we noted that when $\nu \lsim -0.9$ the 
Yukawa couplings of the fermions became too large, 
hence for we display the range of $\nu$ in Fig.~2 to be between $-0.3$ 
and $-1$.

The first result to notice is the value of $\Delta a_\mu^A$ corresponding to 
the essentially $\nu$-independent horizontal dotted curve. This $\nu$ 
independence is only approximate and results from the scale of the figure as 
$\Delta a_\mu^A$ varies by a factor of 
order a few as $\nu$ increases up to 2. This curve lies 
about an order of magnitude below the current experimental $(g-2)_\mu$ bound 
implying that we should obtain no reasonable constraint on the KK masses 
arising from this contribution unless there is a 
substantial change in both the experimental 
central value and the error. The rising dotted line 
is the Higgs boson contribution which we note is rather small over most of the 
interesting range of $\nu$. Scaling the curve to allow 
for the first KK masses to only be as 
large as a few to 10 TeV, we find that the region to the left of 
$\nu \simeq -0.70$ is excluded. 
This rules out a reasonable fraction, $\simeq 30\%$, of the preferred allowed  
range of $\nu$ remaining subsequent to our last analysis{\cite {dhr}}. 

There are however two other points to consider. First, a quick examination 
of the perturbative bound on the Yukawa couplings in the Higgs graph shown in 
Fig.~1 also yields a constraint. Imposing the weak requirement that 
$C_H^2=C_{0i}^{ffH}C_{0j}^{ffH}<16\pi^2$ we obtain the dash-dotted line in 
Fig.~2 at $\nu \simeq -0.77$ excluding the region to its left. A second, 
perturbative-like bound can 
also be extracted from the Higgs loop in Fig.1 when we remove the photon line 
and consider the resulting mass renormalization contribution{\cite {chiv}}. 
We next demand the no fine-tuning requirement 
that the finite part of this graph be not much larger than $m_\mu$; from 
this we explicitly obtain 
the constraint that $C_H^2m_{ij}/16\pi^2$ be not much greater than $m_\mu$. 
This then excludes the region to the left of the solid line at 
$\nu \simeq -0.64$ and results in a stronger bound than that obtained from 
the present experimental 
value of $(g-2)_\mu$. Of course one could repeat this 
exercise for the case of the top quark {\it provided} the value of $\nu$ is 
universal. In this case, the value of $C_H^2m_{ij}$ is 
increased by the factor  $(m_t/m_\mu)^2$ and the resulting bound is 
drastically strengthened. We find that the region to the left of 
$\nu \simeq -0.44$ would now be excluded by this analysis; these 
considerations now exclude more than $\simeq 75\%$ of the previously 
preferred range of 
$\nu$ and leaves only the relatively narrow window between $-0.30$ and $-0.44$ 
as allowed. This would seem to greatly disfavor the possibility of the SM 
being in the RS bulk in the case of a universal mass parameter, $\nu$.

In conclusion the present experimental measurements of $(g-2)_\mu$ do not 
place significant constraints on localized gravity models with the SM field 
content in the bulk for most of the allowed range of $\nu$. 
However, requiring that one loop corrections to the 
fermion masses be of the same order as the measured fermion mass, so that no 
fine tuning of the parameters in the Lagrangian are necessary, severly 
constrains the allowed bulk mass parameter space. This result is obtained 
assuming that the value of $\nu$ is flavor independent and that the 5-d Yukawa 
couplings of the fermions are hierarchical. Given these assumptions, however, 
placing the SM field content in the bulk is tightly constrained by the above 
considerations.

\noindent{\Large\bf Acknowledgements}

The authors would like to thank S. Brodsky, S. Chivukula, Y. Grossman, 
J. Ng, M. Peskin, A. Pomarol, M. Schmaltz, and J.Wells for discussions 
related to this work.

\vskip0.5in

%
\def\MPL #1 #2 #3 {Mod. Phys. Lett. {\bf#1},\ #2 (#3)}
\def\NPB #1 #2 #3 {Nucl. Phys. {\bf#1},\ #2 (#3)}
\def\PLB #1 #2 #3 {Phys. Lett. {\bf#1},\ #2 (#3)}
\def\PR #1 #2 #3 {Phys. Rep. {\bf#1},\ #2 (#3)}
\def\PRD #1 #2 #3 {Phys. Rev. {\bf#1},\ #2 (#3)}
\def\PRL #1 #2 #3 {Phys. Rev. Lett. {\bf#1},\ #2 (#3)}
\def\RMP #1 #2 #3 {Rev. Mod. Phys. {\bf#1},\ #2 (#3)}
\def\NIM #1 #2 #3 {Nuc. Inst. Meth. {\bf#1},\ #2 (#3)}
\def\ZPC #1 #2 #3 {Z. Phys. {\bf#1},\ #2 (#3)}
\def\EJPC #1 #2 #3 {E. Phys. J. {\bf#1},\ #2 (#3)}
\def\IJMP #1 #2 #3 {Int. J. Mod. Phys. {\bf#1},\ #2 (#3)}


\begin{thebibliography}{99}
\bibitem{chiv}
We would like to thank S. Chivukula for pointing out this constraint to us.
\end{thebibliography}
\end{document}